\documentclass[twocolumn]{aa} 
\usepackage{graphicx}

\def\lesssim{\mathrel{\hbox{\rlap{\hbox{\lower3pt\hbox{$\sim$}}}{\raise2pt\hbox{$<$}}}}}
\def\gtrsim{\mathrel{\hbox{\rlap{\hbox{\lower3pt\hbox{$\sim$}}}{\raise2pt\hbox{$>$}}}}}

\begin{document}

\title{A Search for Period Changes in $\delta$ Scuti Stars with the Super-LOTIS Sky Patrol System}

\author{C. Blake\inst{1,2}
             \and
         D.W. Fox\inst{1}
             \and
        H.S Park \inst{3}
              \and
	   G.G. Williams \inst{4}}                  

  \offprints{C.Blake}

   \institute{Division of Physics, Mathematics and Astronomy, 105-24, California Institute of Technology, \\Pasadena, CA 91125,    Cblake@princeton.edu
         \and 
 Department of Astrophysical Sciences, Princeton University, Princeton, NJ 08544 
          \and 
  Lawrence Livermore National Laboratory, 7000 East Avenue, Livermore, CA 94550
	   \and	
 Steward Observatory, University of Arizona, 933 North Cherry Avenue, Tucson, AZ 85721. }

\date{Received 17 September 2002}
\maketitle 
\begin{abstract} 

    We have observed a sample of
    $\delta$ Scuti stars discovered by the ROTSE collaboration in 1999
    with Super-LOTIS in order to characterize changes in their
    pulsation periods over a time baseline of roughly three
    years. Achieving these goals required the creation of an automated
    astrometric and photometric data reduction pipeline for the
    Super-LOTIS camera. Applying this pipeline to data from a June
    2002 observing campaign, we detect pulsations in 18 objects, and
    find that in two cases the periods have changed significantly over
    the three years between the ROTSE and Super-LOTIS
    observations. Since theory predicts that evolutionary period
    changes should be quite small, sources of non-evolutionary period
    changes due to the interactions of pulsations modes are
    discussed. 

  \keywords{}
\end{abstract}

%

\section{Introduction}
    It  is estimated that $\approx$50$\%$  of stars located in the lower
    classical   instability      strip  exhibit     $\delta$     Scuti
    pulsations. Most pulsate in a number of non-radial p-modes, others
    in fundamental and/or overtone  radial modes, and some  possibly
    in  g-modes.   The periods of   these stars are  generally between
    0.014d and  0.25d,    with  amplitudes   up  to    1$\stackrel{\rm
    m}{\textstyle  .}$0. There are a number   of distinct subgroups of
    the  $\delta$ Scuti  pulsating  stars.  In all   cases it is   the
    standard $\kappa$ opacity mechanism  which excites the pulsations,
    but the stars located in this region of  the instability strip may
    be Population I, evolved Population II,  or massive stars evolving
    through the instability strip.  A very thorough description of the
    characteristics of these stars, and their different subclasses, is
    given by Breger (\cite{breger00}).

    The radial pulsation of  a $\delta$ Scuti star  is related to  the
stellar  density  through    the    simple equation   of     pulsation
$P\sqrt{\rho}=Q$. Observed  changes  in the  pulsation  frequencies of
these  stars  therefore provide important   information about  stellar
structure. Since  stellar mass is conserved on  the  time scale of our
observations, finding   period  changes  provides   information  about
changing  stellar radius. Detailed  models of these changes in stellar
structure, and the resulting changes in pulsation frequency, have been
carried out by  Breger $\&$ Pamyatnykh (\cite{breger98}) who predicted
that  pulsation frequencies should be  stable  or slowly increasing as
some  of  these types  of stars evolve   away  from the Zero  Age Main
Sequence (ZAMS). The theoretically   expected exceptions would be  the
rare pre-MS pulsators, of which very few are known, and the Population
II stars. Breger $\&$ Pamyatnykh  (\cite{breger98}) point out that  in
practice what is  often observed is not a  slow evolution of pulsation
frequency, but  abrupt changes more drastic  than can  be explained by
the current models. While Szeidl (\cite{sz00}) provides examples where
period  changes  have been proven  false  by careful re-examination of
data,    many   examples of  well    established   period changes also
exist. These  observed changes  are    not necessarily even   in   the
predicted sense, i.e. some are of increasing  frequency, and there are
examples from the literature where the period  changes are an order of
magnitude   larger  than   expected.   The  predicted,   and sometimes
observed, period changes in  $\delta$ Scuti stars are  generally given
in units of $P^{-1}dP/dt\approx10^{-7}y^{-1}$,  an effect on the order
of   $10^{-8}$  over the  3  year  baseline studied here.  Breger $\&$
Pamyatnykh  (\cite{breger98})   point  out   that  the observed period
changes in Population II stars are often abrupt and  up to an order of
magnitude larger than those in Population  I stars and that stars just
below the Main  Sequence may demonstrate changes  up to two  orders of
magnitude larger than Population I stars.

   A small   gradient  in the  period  of pulsation   requires  a long
   baseline of observations of  at   least a  decade to  detect   with
   certainty. For example, several decades of data for the star EH Lib
   has  been   analyzed  by  a number    of  groups (Mahdy  \&  Szeidl
   (\cite{mahdy80}),Yang  et  al.  (\cite{yang92}),   and  Agerer   \&
   Huebscher (\cite{ager97})) who  reached different conclusions as to
   subtle changes in the period  of pulsation of this star. Obviously,
   systematic effects,  and  effects related  to  the method   of data
   analysis, can be problematic when trying to  identify such a subtle
   change. Light travel time effects due to  binarity, and errors with
   the  O-C analysis, or  time of  maxima  counting, may lead to  such
   ambiguities in  the changing pulsation  period. Here, large changes
   in periods of pulsation  over  a comparatively short,  three  year,
   time   scale are investigated in order    to compare to theoretical
   predictions.


\section{Sample Stars}
A sample  of 18 $\delta$  Scuti stars  was  selected from  the on-line
database of the   Robotic Optical Transient  Search Experiment (ROTSE)
project. The  ROTSE  team  discovered  many  new variable  stars,   as
describe by  Akerlof et  al. (\cite{aker00}), and  has made   the data
publicly
available\footnote{http://umaxp1.physics.lsa.umich.edu/$\sim$mackay/
rsv1/rsv1$\_$home.htm}. Stars from the  ROTSE database were chosen, as
opposed to  stars from the standard catalogs  of $\delta$ Scuti stars,
because the  ROTSE stars were readily observable  at the time  of year
when the instrument  for this study was  available, and were generally
fainter than other samples of $\delta$ Scuti stars, thus better suited
to observation by our instrument. Several months of well sampled ROTSE
observations  could be easily retrieved  for each star from the online
database. Since   the ROTSE  data    were  taken without  a   standard
photometric filter, the  standard $\delta$ Scuti subclass definitions,
based  on  V band photometric  amplitudes, are   difficult to apply. A
selection was made for amplitude $<$ 0$\stackrel{\rm m}{\textstyle.}$4
and  $P<0.25$d,  and only stars  with  the  highest ROTSE  light curve
quality designation were considered. The stars selected from the ROTSE
database could possibly fall into four categories: (1) Population I p-
and/or g-mode pulsators,  (2)  Population I radial fundamental  and/or
overtone  pulsators, generally   called  High  Amplitude  Delta Scutis
because  they have larger   photometric amplitudes, (3)  Population II
stars, generally  called SX  Phe  stars, or  (4)  pre-MS and   evolved
massive  stars. The sample  stars and their  properties as reported in
the ROTSE database are given in Table 1. 

\begin{table*}
    \begin{center}
       \caption[]{The $\delta$ Scuti Star Sample. The listed periods are those determined by the ROTSE team and reported in the the ROTSE database.}
        \label{prop}
         \begin{tabular}{lccccc}
            \hline
            \noalign{\smallskip}
            Star$^{1}$  & RA$\degr$(J2000) & DEC$\degr$(J2000) & Mag &  P(d) & $\Delta$Mag\\  
            \noalign{\smallskip} 
            \hline
            \hline
            \noalign{\smallskip}
                ROTSE 9470   & 253.90332  & 52.379080  & 13.40  & 0.1613 &  0.39 \\
                ROTSE 4861   & 264.90536  & 50.200581  & 12.45  & 0.1761 &  0.35 \\
                ROTSE 3956   & 194.17738  & 23.152025  & 13.54  & 0.1654 &  0.33 \\
                ROTSE 3554   & 205.30734  & 31.790440  & 13.03  & 0.1318 &  0.31 \\
                ROTSE 1526   & 210.27316  & 24.704525  & 12.24  & 0.1939 &  0.10 \\
                ROTSE 3124   & 214.22528  & 23.878872  & 13.06  & 0.1772 &  0.21 \\
                ROTSE 3709   & 215.48952  & 23.449043  & 13.35  & 0.1848 &  0.14 \\
		ROTSE 7388   & 217.66783  & 27.224258  & 14.23  & 0.1744 &  0.29 \\
                ROTSE 4837   & 221.33072  & 35.466972  & 13.77  & 0.1707 &  0.30 \\
                ROTSE 706$^{a}$ & 231.02898  & 36.866928  & 10.99  & 0.1041 &  0.33 \\
                ROTSE 9134   & 232.84311  & 35.881794  & 14.32  & 0.1784 &  0.46 \\
                ROTSE 917    & 243.83421  & 35.707275  & 11.10  & 0.1805 &  0.28 \\
                ROTSE 2573   & 240.62502  & 37.560284  & 12.31  & 0.1794 &  0.10 \\
                ROTSE 3232   & 242.53889  & 35.958557  & 12.68  & 0.1594 &  0.23 \\
                ROTSE 2275   & 243.61665  & 30.529236  & 12.25  & 0.1769 &  0.10 \\
                ROTSE 2592   & 237.44629  & 23.915735  & 12.45  & 0.1110 &  0.12 \\
                ROTSE 2926   & 242.08601  & 28.212255  & 12.66  & 0.1750 &  0.11 \\
                ROTSE 2986   & 240.19507  & 24.261307  & 12.59  & 0.1965 &  0.13 \\
                 
             \noalign{\smallskip}            
              \hline
              \end{tabular}                    
       \end{center} 
        (1) These data were collected from the ROTSE variable star database, available at \\
            http://umaxp1.physics.lsa.umich.edu/$\sim$mackay/rsv1/rsv1$\_$home.htm \\
        ($a$) The known variable star YZ Boo
    
\end{table*}


\section{Observations}
    Observations of the target  stars were carried  out during May and
    June,  2002, with the   Super Livermore Optical  Transient Imaging
    System (Super-LOTIS)  robotic telescope   located at  the  Steward
    Observatory site  on Kitt Peak, Arizona.   The telescope system is
    described   by Park  et  al.(\cite{park97}  and  \cite{park02}). A
    $2048^2$  pixel CCD detector is located  at prime focus  of a 60cm
    f/3.5 Boller \& Chivens Newtonian reflector,  resulting in a field
    of    view   of 0.8$\degr^{2}$     and     a   plate   scale    of
    1.5$\arcsec/$pixel. Typical   seeing at  the Super-LOTIS site  was
    $2\arcsec-3\arcsec$,  so   the   stellar   PSFs   were   generally
    under-sampled. The system  functions autonomously, with a  list of
    targets  being  submitted each day, and  is  capable of collecting
    $\approx$600 individual   30 second  images  during an  eight hour
    observing run. Since   one  of  the  primary  functions  of   this
    instrument is to search for the prompt optical emission associated
    with gamma ray  bursts,  it is able  to   slew very quickly if   a
    satellite trigger  is received. As  a result, the  observations of
    variable stars were occasionally interrupted to chase a burst.

     Images of  the $\delta$ Scuti stars, taken  with  a Bessel R band
     filter,  were  typically 20 to  30   seconds in length,  with two
     images  of   the same field  taken  consecutively.  Each star was
     visited typically six times  per  night, resulting  in 10 to   20
     individual   data points   for  each   star  for each   observing
     night. Since  the Super-LOTIS CCD  camera is  thermo-electrically
     cooled, there is a   non-negligible  dark current which  must  be
     removed with a  dark image. More  than 50 dark  images were taken
     each night, several for each unique exposure duration. At several
     points  during each night a set  of exposures of the random field
     at Zenith was taken for the purpose of constructing a
     super  sky flat by median  filtering  the images of the  randomly
     distributed stars.  While useful for  the study of the periods of
     $\delta$ Scuti stars, these data  were also useful for a thorough
     characterization of   the Super-LOTIS instrument.  Since  the CCD
     camera is  located at prime  focus of  a  fast Newtonian  optical
     system, problems with   focus stability and optical  aberrations,
     such as coma, were common. Mechanical malfunctions with the drive
     system  and  wind  induced vibration,  leading   to poorly guided
     images,  as well as  electronic problems with  the CCD leading to
     noisy images, were also identified. The  12-bit CCD camera itself
     has  seen many years of use,  and will be  replaced with a device
     with   higher  quantum    efficiency    and  better  read   noise
     characteristics. The current CCD has a high gain of $20e^{-}/ADU$
     and an approximate read noise of  $70e^{-}$. While the data still
     produced good results despite these hardware problems, the system
     is constantly being improved. Less data  loss, and higher quality
     images,  may   be expected with     the installation of  a  newly
     purchased  CCD camera  and improvements to   the mount which will
     occur during 2003.

\section{Data Reduction}
    A  pipeline  for the   reduction of  images from  the  Super-LOTIS
    telescope was devised for  use with this  data and, hopefully, for
    use with future projects as well. This pipeline was based on codes
    written in Interactive Data  Language(IDL), and includes code from
    public       IDL      libraries,    such      as    the    Goddard
    library\footnote{http://idlastro.gsfc.nasa.gov/homepage.html}. The
    process was broken  up     into three steps:    image   reduction,
    astrometry, and  photometry.    First, master   dark  images  were
    created.  The dark current was typically measured to be 0.2 ADU/s,
    on top of a bias level of 182 ADU.  The  original intention was to
    make flat-field images by a median filter  of images of the random
    star fields taken throughout the  night at $90\degr$ altitude.  In
    practice, it was determined  that with the  exposure time that had
    been allotted   there were  not  enough sky   counts to produce  a
    statistically meaningful flat-field.  In  order to investigate the
    effects  of  using unflattened  data, a  number  of  images  taken
    occasionally  by  Super-LOTIS  during   the  morning and   evening
    twilight were carefully studied. Since the  CCD camera was removed
    for  maintenance  on several   occasions,  it was  decided  not to
    attempt  to apply  these few  test  flats  to data  taken on other
    nights. For  objects  within the  central  $90\%$ of the CCD  chip
    area, the contribution   to photometric errors   from the lack  of
    flattening is expected  to be $<1\%$.  This is taken into  account
    during photometric calibration when only stars  near the center of
    the  CCD chip  are  utilized for  determination of the photometric
    zero point of each image. 

    A significant portion of the data, up  to $25\%$, were found to be
    unusable for   some    reason  related to   hardware   or  weather
    conditions.  A series of  tests were implemented in order to reject
    these poor images  at the very onset  of processing. A requirement
    was  set  that   there be at    least 100  $5\sigma$  point source
    detections in each image. Since  data were often taken under patchy
    clouds,  this  helped  to  remove images taken   during  very poor
    transparency or  those taken too far into  the morning twilight. A
    test was run to check  the gradient of  the  sky value across  the
    chip from edge to edge,  and a  cut set at   a gradient less  than
    $10\%$.  This helped  to identify  images taken  too close  to the
    moon. Similarly, a test was  run to determine  the gradient of the
    sky values in a set of annuli moving  from the center of the image
    toward the edge. This test easily identified a distinctive problem
    caused by the   occasional   shutter malfunction. Lastly,     poor
    tracking was identified by calculating the autocorrelation of each
    image.   The    characteristic 'hiccup'  caused    by  a  tracking
    irregularity  in the  Right    Ascension gear or  mount  vibration
    induced   by wind, which     is  easily  identifiable  by   visual
    inspection, was identified in the  data pipeline as a strong image
    autocorrelation well  outside  the  normal image  seeing  disk  of
    $\approx5$ pixels. The  autocorrelation was also  found useful for
    identifying the occasional image with a severely distorted PSF due
    to poor focus. This test was done using FFTs of the images and was
    found to be fast enough as to require only a very small portion of
    the total data pipeline CPU time.

    Once  poor quality data were  rejected,  a World Coordinate System
    (WCS) astrometric  solution was added to the  FITS  header of each
    remaining file. Approximate center-of-field coordinates are logged
    by   the telescope control system  at  the time  the exposures are
    taken, so these estimates  were used as  a  starting point for  an
    accurate WCS solution. The pointing  of the Super-LOTIS system was
    typically better than  $6\arcmin$, with an average pointing  error
    of $5.7\arcmin$ over 1000 pointings. Since the field of view is so
    large, targets were seldom,  if ever, not placed  on the CCD chip,
    but  the pointing   accuracy is  expected  to  improve with future
    upgrades to the   telescope mount. The astrometric  solutions were
    determined  by cross correlating   the positions of point  sources
    within  each image with the positions  of stars in  the Guide Star
    Catalog II. Point sources  in the images  were identified  with an
    IDL adaptation of  the Stetson (\cite{stet87}) DAOPHOT  algorithms
    written   by W.Landsman   and   available  in    the  Goddard  IDL
    library. There were  typically 200-300 matches between image stars
    and catalog  stars for  each  image. The cross correlation  of the
    catalog and source positions was maximized allowing translation of
    the image point source positions,  a  rotation of up to  $5\degr$,
    and   a  low order polynomial  term  in  the radial direction. The
    resulting   coefficients   were stored   in    a GSSS  astrometric
    structure, of the type used for  the HST guide star survey images,
    and  added to the image FITS  header. The typical residuals of the
    astrometric fits were $\leq0.5$ pixel. Since the stellar PSFs were
    under-sampled, the accuracy  of the fits  was not as high as would
    be expected   from the application  of this  method  to other data
    sets, for  example Hogg et al.  (\cite{hogg01}), which is followed
    closely here. Parts of the  Super-LOTIS pipeline astrometric  code
    were drawn from the SDSS  PT telescope data pipeline described  by
    Hogg et   al.  (\cite{hogg01}),  but  here   the requirements  for
    astrometric     precision   are   much      less        stringent,
    $\approx1$pixel. The failure rate for the astrometric solutions to
    images passing the image quality tests was less than $10\%$.

    Since the Super-LOTIS images   were not  crowded with stars,   the
    reduced images were  photometered using simple aperture photometry
    with  a radius, matched to  the average seeing, of  6 pixels and a
    sky annulus of radii 10 and 18  pixels. These values were determined
    empirically  after testing   a  variety  of  radius and    annulus
    combinations. Sub-pixel  centroiding of  the aperture was  carried
    out prior  to photometering in order  to compensate for  errors in
    the conversion  from  Right  Ascension and  Declination   to pixel
    coordinates, which were found to be on the order of 1 pixel. Since
    the  stellar PSFs were often very  distorted, an accurate centroid
    was   very  important.  This was   done   using  a  center-of-mass
    estimation.   In each   data  image,  every  star  amounting to  a
    $5\sigma$  detection  above the noise  in  the sky  background was
    photometered and  its   flux,   chip position, Right    Ascension,
    Declination,  and  nearby sky value were  stored,   along with the
    Julian date of the observation, in a FITS  binary table written to
    accompany each data image. The fluxes were converted to magnitudes
    by   comparison  to the  Guide Star   Catalog   II (GSCII) on-line
    database\footnote{http://www-gsss.stsci.edu/support/data$\_$access.htm}. 
The
    stars photometered  in each image were matched  to  stars from the
    GSCII catalog and their   F band photographic  catalog  magnitudes
    recorded. Both  F band and  R band are  red bandpasses, and though
    not identical, are very   similar. While each  individual  catalog
    magnitude   has   a reported   error   of   up to  0$\stackrel{\rm
    m}{\textstyle.}$5, a minimum of 100 of  these values were averaged
    together   in  order   to   substantially  reduce errors  in   the
    photometric zero    points. The   GSCII  contains   more  accurate
    photometry  for bright stars  taken from other  catalogs. For each
    matched star,  a rough zero point was   calculated with the simple
    formula

\begin{equation}
   m_{J_{p}}=-2.5*\log(Flux)+zpt
\end{equation}
\vskip 2mm 
where $m_{J_{p}}$ is the  catalog magnitude from the GSCII,
$Flux$    is the  observed   flux,  in ADU s$^{-1}$,   and $zpt$   is the rough
photometric   zero point. At  least 100   such rough  zero points were
averaged together to create  a final zero point  for each image. Prior
to averaging,  the rough zero  points from the brightest  and faintest
$10\%$ of the stars were removed, and the final average was calculated
on  the remaining values with $3\sigma$  clipping. The errors in these
catalog-based magnitudes are   dominated  by the errors  in  the GSCII
magnitudes, but were found to be accurate to $\approx5\%$.

   Since  in  many situations  accurate  relative  photometry is  more
   important  than  absolute  photometry, the instrumental  magnitudes
   were  further  refined following the   methods of Everett \& Howell
   (\cite{ever01}).  A   small     set   of   stars   brighter    than
   14$\stackrel{\rm m}{\textstyle.}$0, and close  enough to the target
   star that,  based on  known pointing  accuracy  statistics,  all of
   these stars would  appear in each exposure, were  chosen to  act as
   the    reference     for  the      calculation    of   differential
   magnitudes. Assuming  that   each   of the  reference  stars   were
   intrinsically constant in  brightness, small corrections were  made
   to the instrumental magnitudes following the reference stars method
   of Everett \& Howell  (\cite{ever01}, eqn.(2)).  In this method the
   differential magnitudes  are  calibrated using an  average over $N$
   reference    stars weighted  by  the theoretical   variance of each
   observation, in magnitudes.   This estimate was  based on  the read
   noise of the CCD, the photon noise, dark current,  as well as noise
   associated with  the quantization of  the  electrons into  ADU. The
   equation   for $\sigma_{*}$   is   taken from  Everett  \&   Howell
   (\cite{ever01},  eqn.(1)),   but   terms   for   dark  current  and
   quantization  are   added according   to   Howell    (\cite{how00},
   eqn.(4.4))

\begin{equation}
$$\sigma_{*}  =1.0857{\sqrt{F_{*}  +  C_{pix}(F_{sky}    +R^2+F_{dark}
+g^{2}*\wp)}\over{F_{*}}}$$
\end{equation}
\vskip 2mm
where $F_{*}$ is the aperture  flux from the  star in $e^{-}$, $g$  is
the CCD gain in $e^{-}/ADU$,  $F_{sky}$ is the average sky  background
value  in  $e^{-}$/pixel, $F_{dark}$  is  the average dark  current in
$e^{-}$/pixel, $R$ is   the CCD  read  noise in  $e^{-}$, $\wp$   is a
constant related to  the   quantization noise as explained   by Howell
(\cite{how00}), and   $C_{pix}$ relates the  number  of  pixels in the
aperture to the number of pixels in the annulus via

\begin{equation}
$$C_{pix} = n_{aper}[1+(n_{aper}/n_{ann})]$$
\end{equation} 
\vskip 2mm

The  accuracy of  the differential   photometry  was compared to   the
theoretically expected variance in  the magnitude estimations  derived
from the equations for $\sigma_{*}$. The results are shown in Fig.1 as
function of magnitude,   with the heavy  line presenting  the expected
level of photometric  accuracy, due largely to  the high read noise of
the CCD.   By plotting the  measured variance  of  a random  sample of
stars,  all assumed  to be  constant, it is   shown that the level  of
photometric  accuracy    achieved is very  close   to  the theoretical
limit. It   was  found   that  for   a star  of    average  brightness
12$\stackrel{\rm   m}{\textstyle.}$0 the differential magnitude values
produced with this method are  accurate to $\approx2\%$. A photometric
calibration where a number of stars in a single  frame are utilized to
produce  differential     magnitudes,  sometimes    called   'ensemble
photometry', has been   shown  to produce highly accurate   results by
several  groups, including  Everett  \&  Howell (\cite{ever01}).  This
method helps to  reduce systematic effects, as well  as effects due to
color   dependent  extinction,  resulting   in accurate   differential
magnitudes. While  the best photometric  accuracy was achieved when an
ensemble  of  stars was used  to  produce differential magnitudes, the
option exists in the pipeline to derive only the less accurate results
from the GSCII  using a larger  number of stars without requiring that
the all the stars be in every  frame. This is a necessary modification
in order     to quickly produce accurate  magnitudes    for targets of
opportunity, such as  gamma ray burst optical flashes,  when a set  of
ensemble stars can not  be selected prior  to the reductions,  a large
error box must be searched,  and photometric  results are required  as
quickly as possible.
    
\section{Period Determination}
With magnitudes for the sample of $\delta$ Scuti stars produced by the
    photometric pipeline, the periods of variation could be determined
    and  compared to the  values found from  the data taken during the
    1999 ROTSE observations. Since  the  data are unevenly sampled,  a
    periodogram method  was selected for  analyzing periodicity in the
    light  curves.     The effectiveness   of the    periodograms  for
    identifying periodicity in data sampled in a manner similar to the
    ROTSE or  Super-LOTIS    data has been   demonstrated many  times,
    including Schwarzenberg-Czerny   (\cite{cz96}) and Eyer  \&  Blake
    (\cite{eb01}).    Here an   implementation   of  the  Lomb-Scargle
    normalized  periodogram,  a technique based on   the  work of Lomb
    (\cite{lomb76}) and Press \& Rybicki (\cite{pr89}), is utilized in
    an IDL code  based on a publicly  available routine written by  J.
    Wilms\footnote{http://astro.uni-tuebingen.de/software/idl/      \\
    aitlib/timing/scargle.pro}.The statistics    of various  types  of
    periodograms are  treated in  great detail by Schwarzenberg-Czerny
    (\cite{cz98}),  in  which a  common   inaccuracy  in the  standard
    implementation of  the  Lomb periodogram  is pointed out.  In  the
    limiting case  where the number of  data points in the time series
    is very large, the  statistical distribution of the periodogram of
    a white  noise   vector is    well approximated  by   a   $\chi^2$
    distribution,  but in  general  the   distribution is actually   a
    $\beta$ function. In cases were the  number of data points is less
    than $\approx200$,  the    $\beta$  function provides  much   more
    accurate false alarm probabilities for  detected signals. The full
    distribution function, and its limiting case, is given by

\begin{equation}
      $$1-P(z) = (1-{2z\over n})^ {{n\over 2}} \to e^{-z}$$
\end{equation}
\vskip 2mm

The Lomb   method was applied   to the publicly available  ROTSE light
curves  in an  attempt to  recover the  periods  for the  sample stars
published  in the ROTSE database.  The  period search mesh  was set so
that   an  average resolution in  period  of  $10^{-7}$  was achieved.
Heliocentric corrections were made  to the times of observations prior
to analysis with the Lomb-Scargle algorithm.  When weighted by errors,
the periods reported by ROTSE and  those determined from analyzing the
ROTSE   data with   the   Lomb-Scargle algorithm   were identical. The
published ROTSE periods were  determined using a cubic B-spline method
described  by Akerlof  et    al.   (\cite{aker94}). This  method    is
computationally  slower  than  the   Lomb  algorithm,  but    is  more
statistically robust for light curves  which are not well approximated
by sinusoids. Since  the light   curves  of $\delta$ Scuti  stars  are
basically sinusoidal, the Lomb method is  thought to be preferable for
this application. An  example of a  periodogram for the star ROTSE 706
is   given in Fig.3.   While the  statistical  properties of  detected
signals have received much  attention, Akerlof et al.  (\cite{aker94})
points out that frequency error  estimation is rarely discussed in the
literature. A simple Monte Carlo  method was utilized here to estimate
the   errors    in the   detected     periods.  Following  Akerlof  et
al. (\cite{aker00},  Fig.3),  the photometric errors   reported in the
ROTSE   database  were  augmented   to  match  the   published   error
distribution as a function of magnitude. Systematic effects were found
to    increase the ROTSE   photometric  errors  from the theoretically
derived  limit  $0.5\%$   at 10$\stackrel{\rm m}{\textstyle   .}$0  to
$\approx2\%$.   For 200 trials a Gaussian  random vector, of sigma set
to  match  the mean  expected photometric   variance of the individual
measurements  was  added to the photometric   data, processed with the
Lomb algorithm,  and   the resulting highest   power  frequencies were
tabulated. The standard deviation of these frequencies was taken to be
the $1\sigma$ error on  the originally determined period. Periods  and
accompanying errors for the ROTSE and Super-LOTIS photometry using the
Lomb algorithm and Monte Carlo error  estimation are given in Table 2.
The significance of the period differences are also given, in units of
the     quadrature   added   errors  for    the     individual  period
estimations. Inspection of our  derived period changes, and comparison
with our Monte Carlo-derived  period errors in Table  2, may appear to
indicate that    our  period errors   are  overestimated (with reduced
chi-squared  of $<<$1  for the  dataset).  However,  we note that  the
Monte Carlo  technique we  use  will, even in the  best circumstances,
inflate  derived   errors by a  factor  of  $\sqrt{2}$  by  adding its
``false'' errors to the intrinsic  errors  of the data.  Moreover, 
in our   numerical   investigations of  the  period-fitting
process, we found that the period uncertainties  are often limited not
by the shape of the  global minimum at  the true period, but rather by
the existence  of nearby local minima  corresponding to quite distinct
period     solutions.    In  this   case,     the traditional Gaussian
characterization of    the    uncertainties is  inadequate,   and    a
conservative approach --  such as we have adopted  here -- will appear
to have overestimated the errors in those cases  where the true global
minima have (in fact) been derived.

\begin{table*}
  \begin{center}
      \caption[]{Lomb-Scargle periods and Monte  Carlo error estimates for the
      ROTSE data and the Super-LOTIS data}
         \label{prop2}
         \begin{tabular}{lccccc}
            \hline
            \noalign{\smallskip}
                &  ROTSE &  & Super-LOTIS \\
             Star  & P(d) & $\sigma$(d) & P(d) & $\sigma$(d) &  $\Delta$P/$\sigma$\\  
            \noalign{\smallskip}
            \hline
            \hline
            \noalign{\smallskip}
                ROTSE 9470   & 0.161347  &  1.1e-5    &  0.161343  &  4.8e-5   &   0.08\\
                ROTSE 4861   & 0.176131  &  6.7e-6    &  0.176242  &  3.1e-5   &   3.50\\
                ROTSE 3956   & 0.1654    &  0.0033    &  0.1654    &  0.0047   &   0.00\\
                ROTSE 3554   & 0.1318    &  0.0022    &  0.1318    &  0.0022   &   0.00\\
                ROTSE 1526   & 0.193927  &  1.0e-5    &  0.1939    &  0.0077   &   0.00\\
                ROTSE 3124   & 0.1772    &  0.0027    &  0.177173  &  9.3e-5   &   0.01\\
                ROTSE 3709   & 0.185     &  0.010     &  0.1850    &  0.0090   &   0.00\\
		ROTSE 7388   & 0.175     &  0.019     &  0.175     &  0.014    &   0.00\\
                ROTSE 4837   & 0.171     &  0.011     &  0.170756  &  6.9e-5   &   0.02\\
                ROTSE 706$^{a}$    & 0.1040889 &  7.0e-7    &  0.1040453 &  3.9e-6   &   11.0 \\
                ROTSE 9134   & 0.151     &  0.037     &  0.151     &  0.013    &   0.00\\
                ROTSE 917$^{}$    & 0.1805    &  0.0056    &  0.180344  &  1.5e-5   &   0.03\\
                ROTSE 2573   & 0.179     &  0.018     &  0.179     &  0.011    &   0.00\\
                ROTSE 3232   & 0.159372  &  7.6e-6    &  0.159     &  0.011    &   0.03\\
                ROTSE 2275   & 0.177     &  0.016     &  0.1768    &  0.0053   &   0.01\\
                ROTSE 2592   & 0.1199    &  0.0025    &  0.114     &  0.015    &   0.40\\
                ROTSE 2926   & 0.175     &  0.012     &  0.185     &  0.059    &   0.17\\
                ROTSE 2986   & 0.196518  &  1.6e-5    &  0.1965    &  0.0093   &   0.05\\
              \noalign{\smallskip}            
            \hline
          \end{tabular}         
       \end{center} 
       ($a$) The known variable star YZ Boo 
 \end{table*}

\section{Results}
As expected, for the  majority of stars  in the $\delta$ Scuti sample,
the principle  periods of variation  were found, within errors, not to
have  changed between the  1999 ROTSE observations and the Super-LOTIS
observations during May and June, 2002. However, two stars, ROTSE 4861
and ROTSE 706, were  found to exhibit statistically  significant period
variations. In the case of ROTSE 4861 a  period increase of ${\Delta P
\over  P}\approx6.3\cdot10^{-4}$   was detected  at    the 3.5$\sigma$
level.   For ROTSE   706  and period   decrease of    ${\Delta P \over
P}\approx1.4\cdot10^{-3}$ was  detected at the 11.0$\sigma$  level. In
Fig.2 the ROTSE and  Super-LOTIS data sets are  combined and phased to
the ROTSE period. It  is clear from the  folded light curves  of ROTSE
706  that there is a   phase lag between the   two data sets which  is
greater than the error in the phase estimation.  At the same time, for
the  majority of the stars in  the sample the  period errors are large
enough  that the phase  uncertainties  over the three years separating
the observations becomes large, $>.25P$, meaning that the phase is not
expected to be   coherent between the  two  data sets. In the  case of
ROTSE 706, where  a period change  was identified and the phase errors
due to period inaccuracies were negligible, an attempt was made to fit
the  ephemeris  from  the ROTSE and   the  Super-LOTIS observations by
adding a  $dP/dt$  term of  the magnitude   of the  determined  period
change. Since these  observed period changes  are large, and therefore
not    evolutionary,  the ephemeris  were    also  fit  to an  abrupt,
step-function  period change. Neither  case  was found the improve the
phase residuals for ROTSE 706. Since the period changes could be quite
complex in nature,  it might be  expected that these two simple models
of period change would fail to adequately fit the data.

\section{Discussion}
The  magnitude  of these period   changes are  much larger  than those
    expected  from  stellar evolution,  with the  models  of Breger \&
    Pamyatnykh (\cite{breger98}) predicting that Population I $\delta$
    Scuti periods in the 0.1d  to 0.2d range are essentially constant,
    $P^{-1}dP/dt  < 10^{-8}  year^{-1}$,   over the  duration   of the
    observations considered   here. For pre-MS  stars the  predictions
    increase by a  factor  of up to  100,  still at least an  order of
    magnitude  smaller than    the period  changes   found  here.  The
    non-evolutionary sources of period changes are discussed by Breger
    \& Pamyatnykh   (\cite{breger98}).  Given  the  amplitude of   the
    observed period  changes and the  short time span over which these
    changes  occurred, it  seems   unlikely  that light   travel  time
    effects,    resulting     from     binarity,     could          be
    responsible. Additionally, previous searches for period changes in
    ROTSE 706 have not turned up effects  of the magnitude found here,
    indicating that  light travel  effects  due to  a periodic  binary
    motion are even    less likely.  An interesting   possibility   is
    outlined in the work of Moskalik (\cite{mask85}) who describes the
    possible  effect    of  the  non-linear interaction     of various
    pulsational modes within the star  that could produce jumps in the
    pulsation period of  a $\delta$ Scuti star  of the amplitude found
    here. The theoretical predictions for period changes in Population
    II  $\delta$ Scuti stars  are   less well  constrained, but it  is
    thought that such changes are  typified  by large jumps in  period
    followed  by relatively   constant  behavior.  These  jumps    are
    observed,   for   example in  the   star   CY Aqr  by   Powell  et
    al. (\cite{pow95}), and  are  usually of the order   $P^{-1}\Delta
    P\approx10^{-6}$.

    Further observations of ROTSE 4861  should be undertaken in  order
    to determine its age and stellar properties. The star ROTSE 706 is
    a  well studied Population  I  High Amplitude  $\delta$ Scuti star
    also  known as  YZ  Boo. Since this   star has a  high photometric
    amplitude  and  relatively short period,  it  has been included in
    many studies of   the properties  of  $\delta$  Scuti  stars.  The
    long-term stability of the period of  pulsation of YZ Boo has been
    studied extensively by Hamdy et   al. (\cite{hamdy86}) who find  a
    period  increase of  $P^{-1}dP/dt=3\cdot10^{-8}year^{-1}$ over the
    last  several  decades. Clearly, our   result  is incongruous with
    their finding.  Since   this star is   a  Population I fundamental
    radial  mode    pulsator,   as  confirmed    by    Pe\~na  et  al.
    (\cite{pen99})  and Rodr\'iguez et  al. (\cite{rod96}), the abrupt
    period changes predicted for Population  II stars are not expected
    to be  present. The most  promising theoretical  explanation for a
    period change  of the magnitude  observed  here in a Population  I
    star is non-linear mode interaction.  Such interactions have  been
    proposed as a possible explanation for the observed period changes
    in   XX Pyx   reported   by Handler   et al.    (\cite{hand98} and
    \cite{hand00}).  Since  $\delta$   Scuti stars  may  have  a large
    number of excited modes of all  types, and probably many more with
    amplitudes   below   current photometric   detection  limits,  the
    calculation of the specific   modes responsible for the  amplitude
    and  period  variations     is  extremely  difficult.   The   mode
    interactions described  by Moskalik  (\cite{mask85}) are  based on
    only three modes, a radial mode interacting with  two stable g- or
    p-modes  of  low order, and   especially considering that the mode
    growth factor,  $\gamma$, is not precisely known  for  many of the
    modes  identified in    XX  Pyx, the  problem of    fitting a mode
    interaction model to the  observations is  not feasible. While  at
    one  time it was  thought that monoperiodic  High Amplitude $\delta$ Scuti    (HADS) did not show any
    evidence for  types   of pulsation other  than  fundamental radial
    mode, or  sometimes an overtone (see  Rodr\'iguez (\cite{rod96})),
    the mode   resonance   explanation  for period changes    requires
    nonradial modes  to be excited in these  stars. The subject of low
    amplitude nonradial  and high order  radial modes in HADS stars is
    discusses by Garrido et al. (\cite{garr96})  who find that diverse
    types  of modes are   often  found in  HADS   stars when  the  the
    photometric data is of high enough quality.  At the same time, the
    mode resonance  hypothesis requires that the  dominant, presumably
    radial,  mode be   unstable,   meaning  nonadiabatic.  Nonadiabtic
    simulations of stellar models  similar to $\delta$ Scuti stars are
    considered  by Stellingwerf  (\cite{stell80})  who finds  that the
    fundamental radial  mode of a stellar  model similar  to YZ Boo is
    unstable with a mode  modulation time scale, $\gamma^{-1}$, of the
    order years. It is certainly plausible that the conditions, namely
    multiple  modes and an  unstable dominant  mode, could  exist in a
    $\delta$ Scuti star, such  as YZ  Boo,  for period changes of  the
    order  $\Delta P /  P  \approx 10^{-4}$  to occur through resonant
    mode  coupling. There   are relatively   few  discussions in   the
    literature of  observations  of large amplitude,  non-evolutionary
    period  changes. In addition to  the case already  discussed of XX
    Pyx, period variations  in V1162 Ori  are presented by Arentoft et
    al.  (\cite{arent01}),  and  the  peculiar  behavior the  one-time
    monoperiodic HADS star RY Lep is discussed by Laney \& Schwendiman
    (\cite{laney02}). In both  cases, changes in  periodicity, similar
    in magnitude to those presented here, are found. As pointed out by
    Laney    \& Schwendiman,  these   abrupt   changes in  period,  or
    ``aperiodic cycles'' may even  be typical of variable stars. Since
    they occur rarely, evidence for such behavior is only now becoming
    more abundant as more data than  ever is being collected by groups
    all over the world interested in $\delta$ Scuti stars.

    Comparatively little is known about ROTSE 4861, the other star for
    which  we  find a period  change.  The  colors  of the  star  were
    measured  with  The 60'' Oscar Meyer Telescope on Mt. Palomar  and  found  to be
    $V-I=1.37$,  almost  a magnitude  more red  than   a typical lower
    instability  strip star. One interesting   note about this star is
    that   its position,  as   reported  in  the  ROTSE  database  and
    re-calculated by us, is located  $5\arcsec$ from the X-Ray  source
    1RXS   J173936.5+501207 from the ROSAT   Bright  Source Catalog of
    Voges   et  al.    (\cite{vog99}). This  raises    the interesting
    possibility that ROTSE 4861 is a new pre-MS, or Herbig Ae/Be, star
    displaying $\delta$ Scuti  variability. Only nine such  stars have
    been identified,  but the discovery, and  extensive study, of such
    objects may prove useful for determining the time-scales
    of pre-MS evolution and  the structure of  young stars (see  Kurtz
    (\cite{kurtz02})). Further observations of ROTSE 4861 are required
    to determine the nature of this star, but Zinnecker $\&$ Preibisch
    (\cite{zinn94}) found that about half   of the Herbig Ae/Be  stars
    they studied were X-Ray bright. If found  to be a pre-MS star, the
    large amplitude and comparative regularity  of ROTSE 4861 may make
    it the best  testbed yet  found for  evolutionary theories. It  is
    predicted  by  Breger $\&$  Pamyatnykh (\cite{breger98})  that for
    pre-MS stars the observed period  derivative is negative, which it
    is  here, but the  amplitude of the change  found here is about 80
    times  larger  than what  is predicted  for  a pre-MS star  with a
    pulsational period of 0.17d.

\section{Conclusions}

   Tested with  imaging data for a set  of 18  $\delta$ Scuti stars, a
   newly  designed  data   reduction pipeline  for  the    Super-LOTIS
   telescope is found to be  efficient and accurate. These photometric
   data were used to  compare  the current  periods of these  stars to
   periods determined  three years   ago   by the ROTSE   project.  As
   expected, the majority of the  stars  show the same periodicity  in
   the ROTSE data and the  Super-LOTIS data. A Monte Carlo  simulation
   was done to estimate the accuracy  of the periods determined with a
   Lomb-Scargle periodogram method,   and   two stars were   found  to
   exhibit  statistically significant period  changes. The  star ROTSE
   706 (=YZ Boo) is a well  studied monoperiodic high-amplitude $\delta$ Scuti which appears to
   have  undergone  a non-evolutionary period  change  during the time
   between the  two data sets  studied here. A   few other examples of
   large  amplitude period changes  in $\delta$ Scuti stars were found
   in the  literature. Along  with  these observed period  changes was
   found discussion of  a possible explanation  through the phenomenon
   of resonant mode coupling   between a nonadiabatic radial mode  and
   two stable nonradial  modes. The second  period change was found in
   the star  ROTSE 4861.  Little is known  about this  star, but it is
   possibly an    X-ray source reported  in   the  ROSAT  BSC. Further
   observations   of this star  are required  to  determine if it is a
   pre-MS  or Herbig Ae/Be  star, in  which case it  would represent a
   prime  target for an  astroseismology  campaign to learn more about
   pre-MS evolution.

   \begin{figure}[t]
        \resizebox{\hsize}{!}{\includegraphics{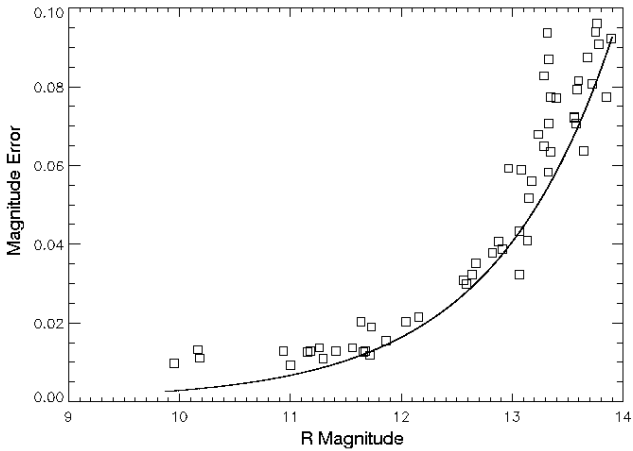}}
        \caption[]{The dark line represents the theoretically expected
        photometric  accuracy based on  equation  (2). The squares are
        measurements   of  the variance in   the  magnitudes of actual
        stars. }
        \label{Fig1} \end{figure}

   \begin{figure}[t] \centering
      
       \resizebox{\hsize}{!}{\includegraphics{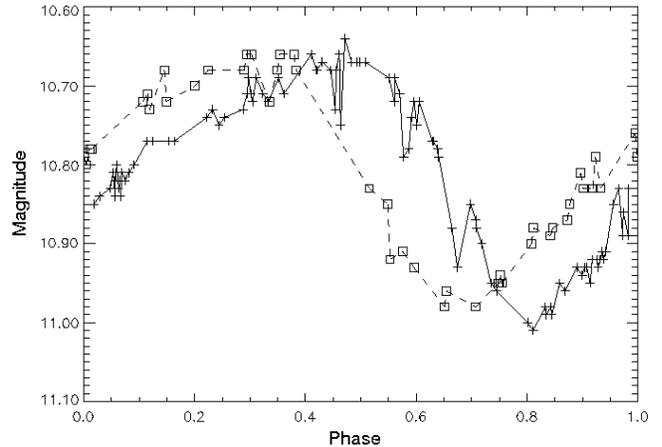}}
        \caption[]{The  ROTSE and  Super-LOTIS data   for  YZ Boo  are
        combined and phased to the  same  period. The solid line  with
        cross points are  the ROTSE data and the  dashed line with box
        points are  the Super-LOTIS data.The statistically significant
        phase lag is clearly shown.}  \label{Fig2} \end{figure}

   \begin{figure}[t]
        \resizebox{\hsize}{!}{\includegraphics{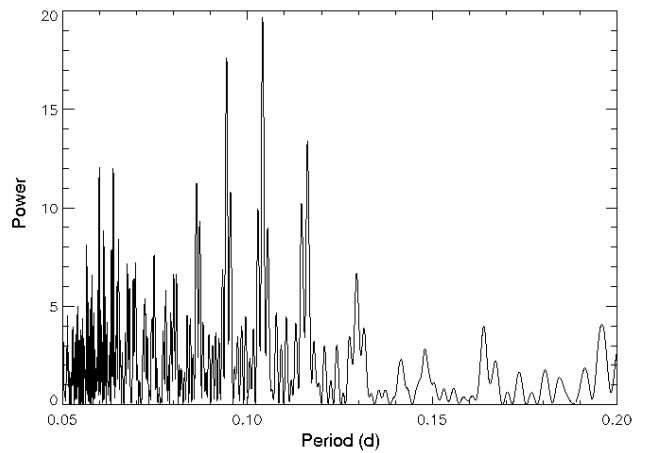}}
        \caption[]{An example of  a Lomb-Scargle periodogram generated
        from   the  Super-LOTIS  observations  of   the star  YZ Boo.}
        \label{Fig3} \end{figure}

\begin{acknowledgements}

   The   authors would like  to acknowledge  Professor Shri Kulkarni's
 crucial role  in   enabling  this collaboration   and his   financial
 support.  The authors would also like to  thank Edo Berger and Alicia
 Soderberg for assistance with  the multi-color observations of  ROTSE
 4861,  and the referee, Torben  Arentoft, for his  careful reading of
 the original manuscript and his detailed  comments.  CB would like to
 acknowledge  the  support of  the  Caltech  SURF  program during  the
 completion of this work.  Support  for this work was provided through
 Professor  Kulkarni   via   NSF grant AST-0098676     and STSCI grant
 HST-GO-09180.01-A  under  NASA  Contract  NAS5-26555.  This  work was
 performed under the auspices of the U.S.  Department of Energy by the
 University   of  California, Lawrence  Livermore  National Laboratory
 under contract No. W-7405-ENG-48  and NASA contract S-03975G.   . The
 Guide  Star Catalog-II  is  a joint  project  of the  Space Telescope
 Science Institute and  the Osservatorio Astronomico di Torino.  Space
 Telescope Science Institute    is  operated by   the   Association of
 Universities for Research in  Astronomy, for the National Aeronautics
 and Space Administration under contract NAS5-26555. The participation
 of the Osservatorio Astronomico di Torino is supported by the Italian
 Council for Research in Astronomy.  Additional support is provided by
 European Southern  Observatory, Space Telescope European Coordinating
 Facility, the International   GEMINI project and  the  European Space
 Agency Astrophysics Division.
     
\end{acknowledgements}

\end{document}